\documentclass{article}
\usepackage{spconf,amsmath,graphicx,amssymb,booktabs,multirow,hyperref,cite}
\usepackage{marvosym}
\usepackage{microtype}
\usepackage{subfigure}
\usepackage{subfigure}
\usepackage{array}
\usepackage{tabularx}
\usepackage{listings}
\usepackage{xcolor}
\usepackage{threeparttable}

\title{SuperCodec: A Neural Speech Codec with Selective Back-Projection Network}
\name{Youqiang  Zheng$^{1}$, Weiping Tu$^{1,2,}$ \textsuperscript{\Letter}, Li Xiao$^{1}$, Xinmeng Xu$^{1}$
\thanks{\Letter Corresponding author: Weiping Tu(tuweiping@whu.edu.cn)}
\thanks{The numerical calculations in this paper have been done on the supercomputing system in the Supercomputing Center of Wuhan University.}}

\address{$^{1}$NERCMS, School of Computer Science, Hubei Luojia Laboratory,\\Wuhan University, Wuhan 430072, China\\
$^{2}$Hubei Key Laboratory of Multimedia and Network Communication Engineering, \\Wuhan University,Wuhan 430072, China\\
}
\begin{document}
%
\maketitle

\begin{abstract}
Neural speech coding is a rapidly developing topic, where state-of-the-art approaches now exhibit superior compression performance than conventional methods. Despite significant progress, existing methods still have limitations in preserving and reconstructing fine details for optimal reconstruction, especially at low bitrates. In this study, we introduce SuperCodec, a neural speech codec that achieves state-of-the-art performance at low bitrates. It employs a novel back projection method with selective feature fusion for augmented representation. Specifically, we propose to use Selective Up-sampling Back Projection (SUBP) and Selective Down-sampling Back Projection (SDBP) modules to replace the standard up- and down-sampling layers at the encoder and decoder, respectively.  Experimental results show that our method outperforms the existing neural speech codecs operating at various bitrates.  Specifically, our proposed method can achieve higher quality reconstructed speech at 1 kbps than Lyra V2 at 3.2 kbps and Encodec at 6 kbps.

\end{abstract}
\section{Introduction}
\label{sec:intro}
Speech coding is essential in modern communications, aiming to compress speech signals to minimal bits with minimal distortion. Traditional techniques like Opus \cite{opus}, Codec2 \cite{rowe2011codec}, and MELP \cite{supplee1997melp} have demonstrated good performance by leveraging psychoacoustics to extract parameters and using codebooks for compression.  However, these traditional codecs have limitations in low-bitrate scenarios due to the inevitable increase of quantization error.

Deep neural networks have significantly improved the state-of-the-art performance of speech coding in two ways. The first one is to replace the synthesizer of the traditional codecs with strong generative models \cite{kleijn2018wavenet, valin19_interspeech, lyra, SSMGAN, zheng2023cqnv} to improve decoded speech quality. For example, Lyra \cite{lyra} is based on an auto-regressive WaveGRU model that synthesizes speech from quantized mel-spectrum, producing high-quality
speech at 3 kbps. The second way is an increasing trend in employing end-to-end coding schemes for speech coding in more recent research works \cite{2021soundstream, pia22_interspeech, jiang2022end, 2022encodec, 2023lmcodec, jiang2023disentangled}.
\begin{figure}[t]
    \centering
    \includegraphics[width=0.45\textwidth]{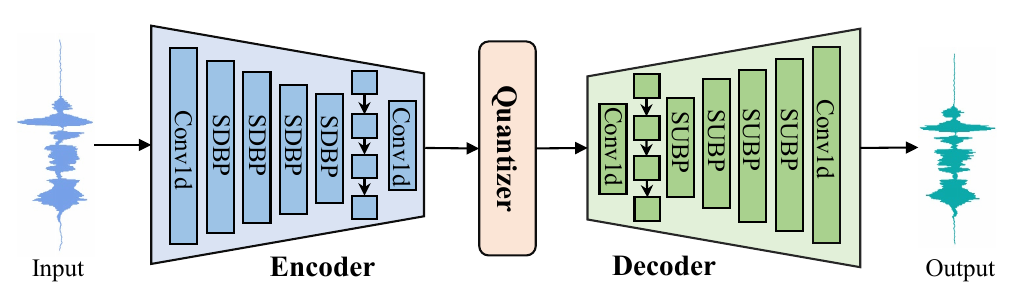}
    \caption{The architecture of SuperCodec.}
    \label{fig:archi}
\end{figure}
These methods utilize the VQ-VAE \cite{van2017vqvae} framework together with a convolutional-based encoder-decoder architecture. Convolutional layers are used in the encoder to down-sample the input speech so as to compress the data, and the transposed convolutional layers are used to reconstruct the speech signal. As a representative example of the end-to-end models, Encodec \cite{2022encodec} at 1.5 kbps demonstrates superior performance compared to Opus \cite{opus} at 6 kbps.


Compared to methods based on generative decoder models, the end-to-end models have significantly improved coding efficiency by achieving high quality at low bitrates. However, existing neural end-to-end speech codecs still encounter limitations in faithfully reconstructing the original speech signal, especially when the bitrate is below 1.5 kbps. Two significant drawbacks can be summarized:
1) \textbf{Missing Information}: Current methods extract the latent representation from the input signal using simple convolutional layers. While proficient at extracting contextualized and non-linear information, these convolution layers face challenges in preserving all information that is used to reconstruct speech at the decoder, while eliminating redundancy in the down-sampling process \cite{xu2023all}.
\begin{figure*}[h]
    \subfigure[SDBP]{\includegraphics[width=0.46\textwidth]{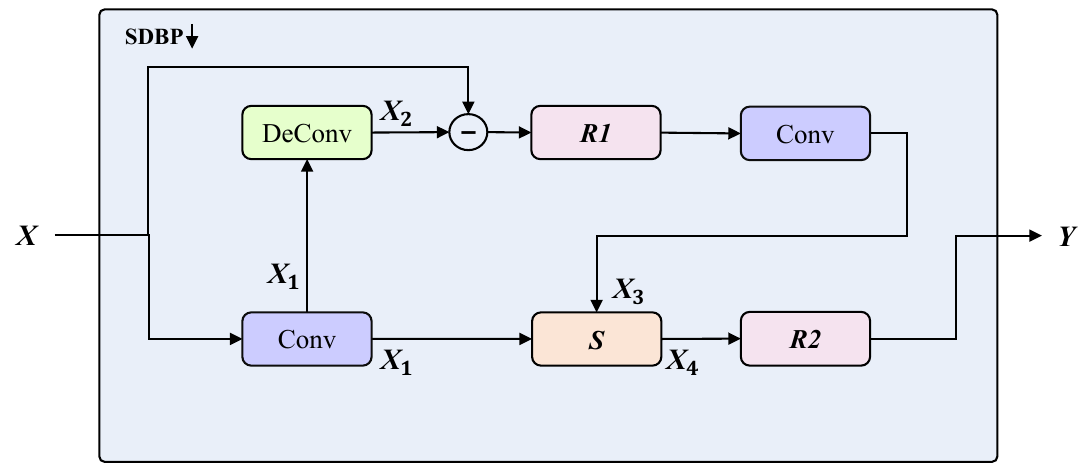}}
    \subfigure[SUBP]{\includegraphics[width=0.46\textwidth]{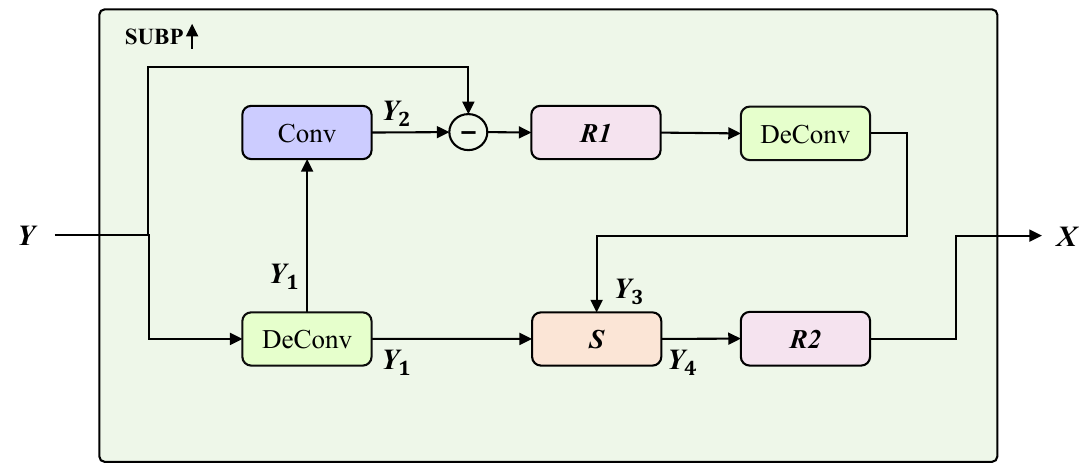}}
    \caption{Network architecture of proposed Selective down and up back-projection. (a) is the selective down-sampling back-projection network(SDBP). (b) is the selective up-sampling back-projection network(SUBP). \textit{\textbf{R1}} and \textit{\textbf{R2}} are the residual blocks, respectively, consisting of the convolution layers with kernel size = 3 and dilation rates = [1, 3], and two convolution layers. \textit{\textbf{S}} is the selective feature fusion network.  }
    \label{sdbp_subp}
\end{figure*} 
2) \textbf{Stumbling in reconstruction}. The low-resolution input representations are upsampled back to the original speech signal using transposed convolutional layers in the decoder. This technique makes it even harder for the neural speech codec to infer fine-grained information for optimal reconstruction.


In this paper, we present SuperCodec, 
a neural speech codec that replaces the standard feedforward up- and down-
sampling layers with Selective Up-sampling Back Projection (SUBP) and Selective Down-sampling Back Projection (SDBP) modules. Our proposed method efficiently preserves the information, on the one hand, and attains rich features from lower to higher layers of the network, on the other. Additionally, we propose a selective feature fusion block in the SUBP and SDBP to consolidate the input feature maps. Our contributions are summarized as follows:
\begin{itemize}
    \item We propose SuperCodec\footnote{Our code is publicly available at:\textcolor{magenta}{\url{https://github.com/exercise-book-yq/Supercodec}}}, a neural speech codec that introduces a novel back projection approach capable of reconstructing high-quality speech signals at low bitrates.
    \item We introduce an effective feature fusion block in the SUBP and SDBP modules, which extracts richer representations to consolidate the input feature maps.
    \item Subjective and objective experiments demonstrate the superiority of our method over existing approaches, even when they use more than 3x the bitrate.
\end{itemize}

\section{Proposed Model}

\label{sec:method}
\subsection{Overall Framework}
Our framework consists of three components: (1) a feature encoder network that maps a raw speech signal $\mathbf{x} \in[-1,1]^T$ of length \textit{T} to a sequence of latent speech representations $\mathbf{e} \in \mathbb{R}^{T_e \times N_e}$, where ${T_e}$ is the length and ${N_e}$ is the dimension; (2)  a residual quantizer searches the corresponding discrete representation of $\mathbf{e}$ with error minimization and its index code in codebooks; (3) a decoder synthesizes the speech signal from the de-quantized representations. Distinguishing the work of existing works, the encoder side consists of four sequential SDBP modules responsible for down-sampling, and the decoder side consists of four sequential SUBP modules responsible for up-sampling, as shown in Fig.\ref{fig:archi}. In our proposed model, the encoder outputs 256-dimensional speech features with a frame rate of 50 Hz from speech at 16 KHz. As for the quantizer, we use residual VQs introduced in \cite{2021soundstream} to transmit continuous speech features over low bitrate. 

\subsection{Selective Back Projection Blocks}
The deep networks of exiting neural codecs commonly use the standard casual convolution and deconvolution layers as the downsampling and upsampling operators to produce lower- and higher-resolution feature maps. However, this mechanism may stumble in preserving details crucial to faithful reconstruction. Back projection iteratively utilizes the feedback residual to refine high-resolution feature maps based on the assumption that the projected, down-sampled version of high-resolution feature maps should be as close to the original low-resolution feature maps as possible. We adopt and extend this technique to solve neural speech codec problems. Specifically, we propose to replace the standard convolution and deconvolution layers with SDBP $\downarrow$ and SUBP $\uparrow$ at the encoder and decoder, respectively. 

As illustrated in Fig.\ref{sdbp_subp}, we utilize the complementary information from back projection to get refined feature maps which in turn produce features of higher quality in the next stage.  It progressively improves the features that propagate throughout the computation. Taking the up-sampling as an example, our SUBP $\uparrow$ module refines the output feature map \textit{${Y}_{4}$}, up-sampled from \textit{${Y}_{1}$} by applying the reverse mapping to recover its original resolution. Despite having the same resolution, the re-sampled feature map \textit{${Y}_{3}$} encloses details that are not previously available to \textit{${Y}_{1}$}. These feature maps are then integrated into \textit{${Y}_{4}$} using a fusion block \textit{S}.

\begin{figure}[h]
    \centering
    \includegraphics[width=0.45\textwidth]{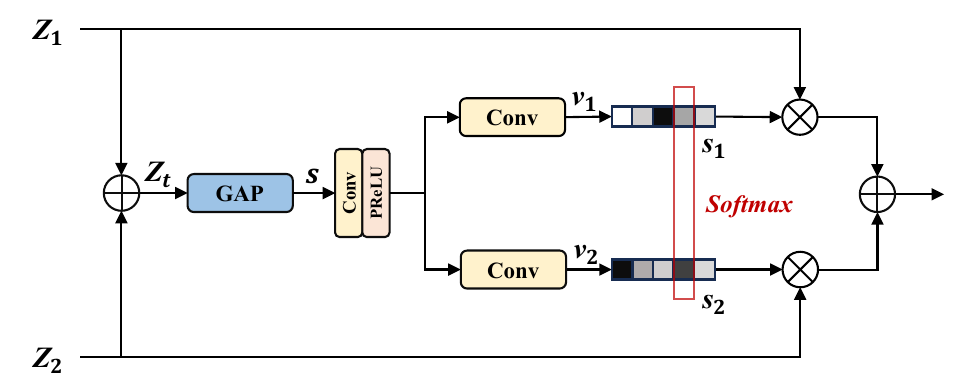}
    \caption{Schematic for selective feature fusion block. It operates on features and performs aggregation based on self-attention. \textbf{GAP} is the Global Average Pooling. $\oplus$ is the element-wise summation and $\otimes$ is the element-wise product operation.  }
    \label{fig:snet}
\end{figure}

\textbf{Selective feature fusion.} The selective feature fusion module performs dynamic adjustment of the respective inputs, as illustrated in Fig.\ref{fig:snet}. Motivated by \cite{skn, xu2023case}, we adaptively aggregate the information from different receptive field features using a self-attention mechanism. This module receives inputs from two parallel features and uses an element-wise sum to combine \textit{${Z}_{1}$} and \textit{${Z}_{2}$}. Then it applies global average pooling (GAP) along the time dimension of \textit{${Z}_{t}$} $\in \mathbb{R}^{N_e \times T_e }$ to compute a statistics \textbf{s} $\in \mathbb{R}^{N_e \times 1}$. Two feature descriptors \textbf{$v_1$} and \textbf{$v_2$}  $\in \mathbb{R}^{N_e \times 1}$ are provided by two parallel convolution layers. And then we apply the softmax function to these descriptors to yield attention activations \textbf{$s_1$} and \textbf{$s_2$} for adaptively recalibrating different feature maps \textit{\textbf{$Z_1$}} and \textit{\textbf{$Z_2$}}. The overall process of feature recalibration and aggregation is defined as Equation \ref{select}.

\begin{equation}
\mathbf{U}=\mathbf{s}_1 \cdot \mathbf{Z}_1+\mathbf{s}_2 \cdot \mathbf{Z}_2
\label{select}
\end{equation}

\subsection{Training Paradigm}

We adopt our framework trained with adversarial loss. The adversarial training framework includes waveform domain and short-time Fourier Transform (STFT) domain discriminators, which follow the Soundstream model \cite{2021soundstream}. We train the SuperCodec model using the standard adversarial loss, feature matching loss following \cite{2021soundstream}.
Furthermore, we use the codebook size $2^{10}$ and vary the number of layers in the RVQ, taking values from the set $\{2, 4, 6, 12\}$, corresponding to 1 kbps, 2 kbps, 3 kbps, and 6 kbps. The adversarial training lasts for about 800k steps.

\section{Experiments}
In the set of experiments, our goal is to validate the effectiveness of our proposed method at different bitrates. We focus on the performance of the SuperCodec model at various bitrates. 

\textbf{Datasets.} The VCTK, a multi-speaker dataset, as described in~\cite{VCTK}, is used to train and evaluate our proposed method.  The total length of the audio clips is approximately 40 hours, and the sample rate of the audio is 44.1 kHz. We downsample the speech data to  16 kHz for training and testing.  Our training set comprises data from 100 speakers, including 57 females and 43 males.  Four female and four male speakers are randomly selected to be employed as unseen speakers condition for the testing.

\textbf{Evaluation Metrics.} We evaluate SuperCodec using both subjective and objective evaluations.
For subjective evaluation, the MUSHRA methodology~\cite{mushra}, with a hidden reference and a low anchor, is used to measure the subjective quality of the reconstructed speech by human raters—a group of twenty listeners, including ten females and ten males, aged between 20 to 27. Twenty utterances, randomly selected from the test set, were evaluated. In addition, Speex~\cite{valin2007speex} at 4 kbps is used as a low anchor. 
As for objective metrics, we following the similar research \cite{2023lmcodec} employ STOI \cite{2010stoi}, ViSQOL~\cite{2020visqol} and WARP-Q~\cite{jassim2021warp} to measure the objective quality of the proposed method.
The sample rate of all data is 16 kHz in our experiments.

\subsection{Quality Evaluation}


We compare our proposed method with existing state-of-the-art neural speech codecs. We use two more mainstream systems as the baselines for our comparison experiments. One is Lyra V2\footnote{\url{https://github.com/google/lyra}}, and the other is Encodec. Specifically, Lyra V2 integrating Soundstream \cite{2021soundstream} gets state-of-the-art coding performance at 3.2 kbps with decreased computational complexity. For Encodec, we retrain Encodec with the same experimental configuration for a fair comparison. We also select the 24 kHz pre-trained model to synthesize speech at 3 kbps and 6 kbps without using  Transformer language model\footnote{\url{https://github.com/facebookresearch/encodec}}. The synthesized signals of the pre-trained Encodec are resampled from 24kHz to 16kHz. The pre-trained Encodec is also tested and the results confirm that our re-trained Encodec is more effective.

\begin{figure}[h]
    \centering
    \includegraphics[width=0.45\textwidth]{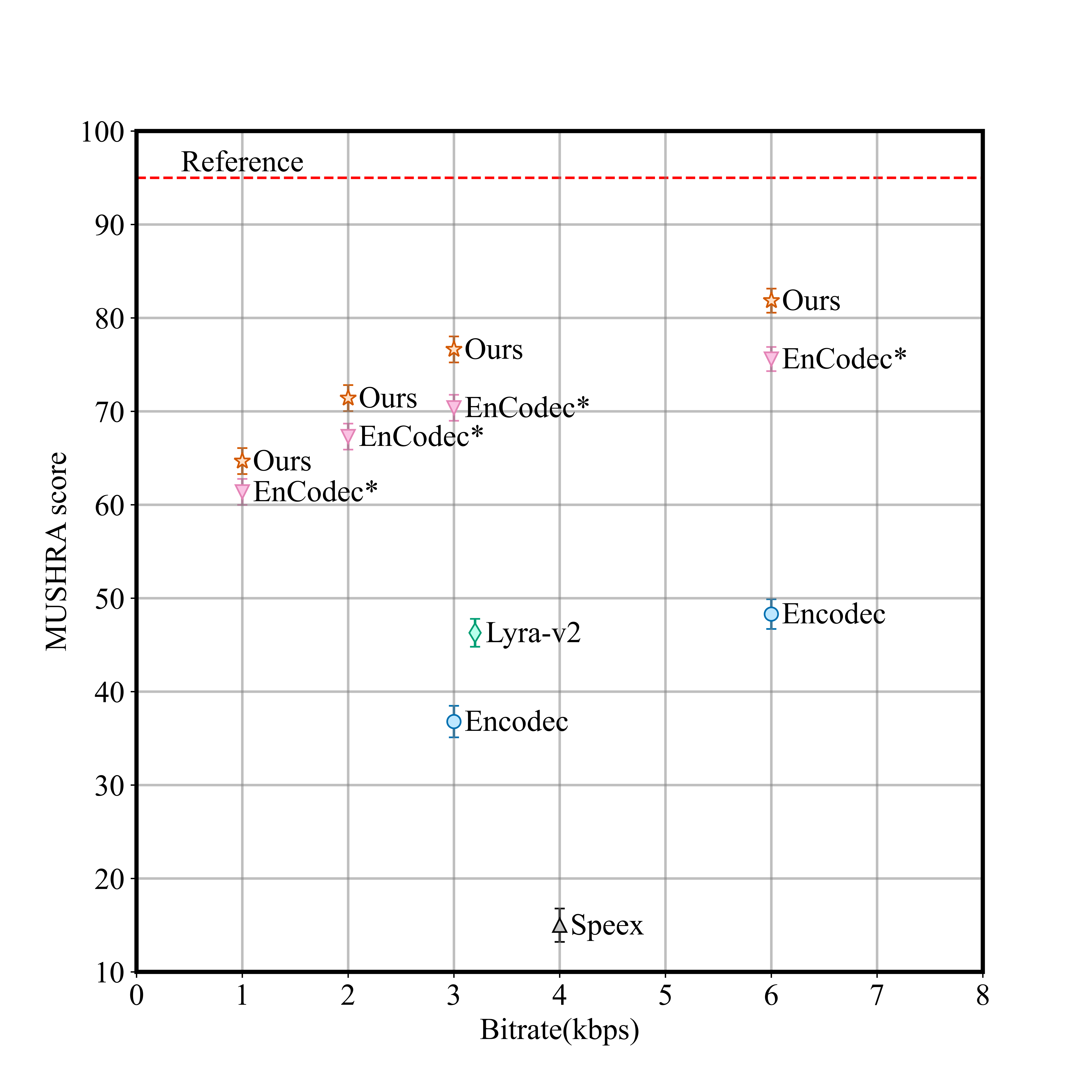}
    \caption{MUSHRA subjective test. The indicated interval in
black represents the 95\% confidence interval for each score.
}
    \label{fig:mushra}
\end{figure}

\begin{figure}[h]
    \centering
    \includegraphics[width=0.48\textwidth]{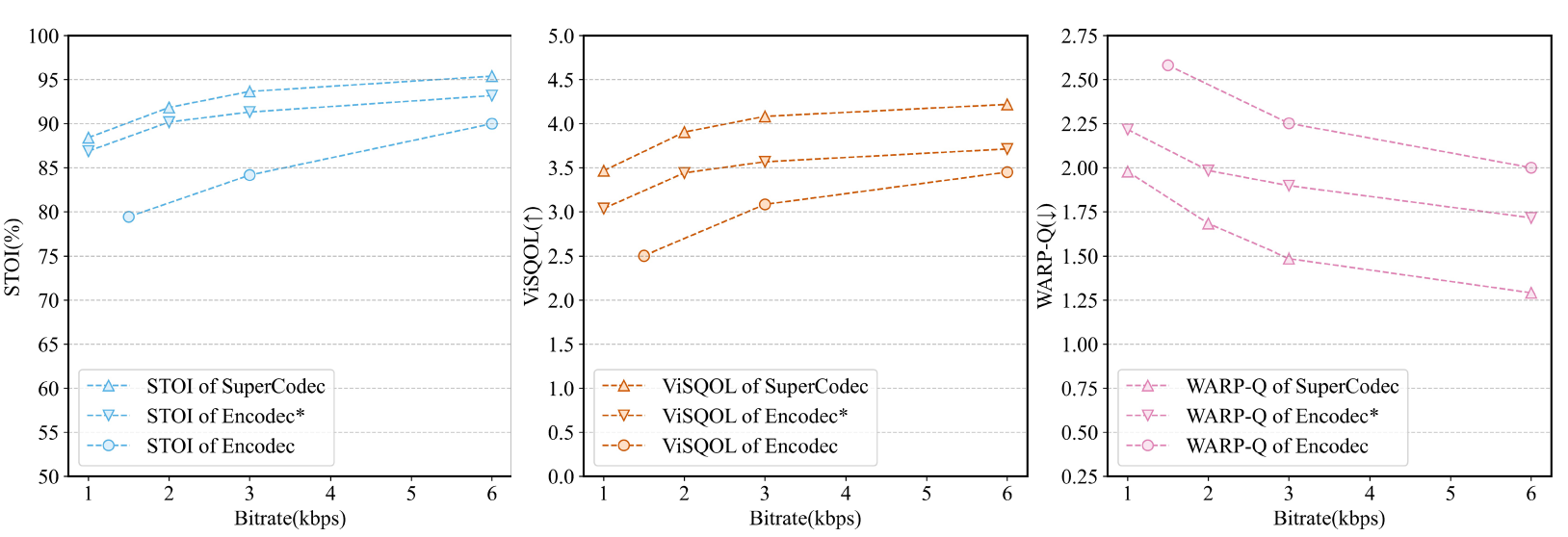}
    \caption{Objective evaluation of SuperCodec at 1 kbps, 2 kbps, 3 kbps, 6 kbps. We compare our method with existing neural speech coding works using STOI, ViSQOL, and WARP-Q.}
    \label{fig:stoi-visqol-warpq}
\end{figure}
 
\textbf{Subjective Results.} As shown in Figure \ref{fig:mushra}, we can see that SuperCodec at 1 kbps outperforms Lyra V2 at 3.2kbps and Encodec 6 kbps. It is also observed that our SuperCodec consistently outperforms the re-trained Encodec at equivalent bitrates, underscoring the superiority of our approach. This similar result persists across different operational bitrates, including 2 kbps, 3 kbps, and 6 kbps. When operating at 6 kbps, SuperCodec gets better performance than all other existing state-of-the-art models. Notably, SuperCodec at 2 kbps surpasses re-trained Encodec at 3 kbps, while at 3 kbps, it outperforms re-trained Encodec at 6 kbps.   These findings firmly establish the effectiveness of the proposed model across a diverse array of bitrate ranges\footnote{Speech samples can be found under the following link: \textcolor{magenta}{\url{https://exercise-book-yq.github.io/SuperCodec-Demo/}}}.

\textbf{Objective Results.} Turning to the objective evaluation, we present it on the speech examples from our test set. As depicted in Figure \ref{fig:stoi-visqol-warpq}, we compare our SuperCodec from 1 kbps to 6 kbps to pre-trained Encodec from 1.5 kbps to 6 kbps and re-trained Encodec from 1 kbps to 6 kbps. When operating at 1 kbps, our SuperCodec significantly outperforms pre-trained Encodec at 6 kbps and re-trained Encodec at 2 kbps according to ViSQOL and WARP-Q. We can obviously observe that SuperCodec gets better performance than re-trained Encodec when operating at the same bitrate. The results further demonstrate the proposed model's effectiveness at low and high bitrates.

\renewcommand{\thefootnote}{\fnsymbol{footnote}}
\footnotetext[1]{We retrain the Encodec model with the same experimental configuration.}
\renewcommand{\thefootnote}{\arabic{footnote}}


\subsection{Ablation Study}
\begin{table}[h]
\centering
\caption{Objective evaluation of SuperCodec at 2 kbps. Ablation studies validate
the effectiveness of \textbf{SDBP} and \textbf{SUBP}.}
\vspace{+1.5mm}
\label{tab:ablation}
 \setlength{\tabcolsep}{1.2mm}{%
\begin{tabular}{lcll}
\toprule
Method               & ViSQOL  & STOI(\%)   & WARP-Q($\downarrow$) \\ \midrule
SuperCodec           & \textbf{3.904}  & \textbf{91.84}  & \textbf{1.683}  \\
SuperCodec w/o.SDBP & 3.847  & 91.28  & 1.720  \\
SuperCodec w/o.SUBP & 3.770  & 90.03  & 1.812  \\
 
\bottomrule
\end{tabular}%
}
\end{table}

\begin{table}[h]
\centering
\caption{Number of parameters and real-time factors for generation on CPU (Intel(R) Xeon(R) Gold 6130H CPU @ 2.10GHz
) and GPU (NVIDIA RTX 3090 GPU)  of SuperCodec at 3 kbps against Encodec~\cite{2022encodec} at 3 kbps on the test dataset with the 24 kHz sampling rate.}
\vspace{+1.5mm}
\label{tab:complexity}
\setlength{\tabcolsep}{1.3mm}{%
\begin{tabular}{llllll}
\toprule[1pt]
\multicolumn{1}{c}{\multirow{2}{*}{Model}} & \multicolumn{1}{c}{\multirow{2}{*}{Parameters ($\downarrow$)}} & \multicolumn{2}{c}{CPU($\downarrow$)} & \multicolumn{2}{c}{GPU($\downarrow$)} \\ \cline{3-6} 
\multicolumn{1}{c}{} & \multicolumn{1}{c}{} & Enc. & Dec. & Enc. & Dec. \\ \hline
Encodec              & 14.85 M      & 0.033    & 0.034    & \textbf{0.004}    & 0.007    \\
SuperCodec             & \textbf{14.66} M              & \textbf{0.030}    & \textbf{0.032}    & 0.005    & \textbf{0.002}    \\ \bottomrule[1pt]
\end{tabular}%
}
\end{table}
We present ablative experiments to analyze the contribution of SDBP and SUBP modules of our model. We measure the STOI, ViSQOL, and WARP-Q on our test dataset. All the ablation experiments are performed for the speech compression task with the same training steps at 2 kbps. Table \ref{tab:ablation} shows that removing  SUBP at the decoder causes the largest performance drop. Replacing SDBP or SUBP with a standard convolution layer yields a  3.43 \% or 1.46 \% decrease in ViSQOL, 1.97 \% or 0.6\% decrease in STOI, and 2.20 \% or 7.66 \% increase in WARP-Q, respectively. We also observe that the SUBP module is more effective than the SDBP module, which validates that the SUBP module we propose is useful for reconstructing the speech.

\subsection{Complexity and Computation Time}
As shown in Table \ref{tab:complexity}, our proposed model has fewer parameters than that of the reference model \cite{2022encodec}. The real-time factor is defined as the ratio between the processing time and the duration of the speech.  An RTF of less than 1 indicates faster than real-time processing. On average, SuperCodec gets better than Encodec in many scenarios except for encoding at GPU, which makes it a good candidate for real-life applications.

\section{Conclusions}
In this paper, we propose a neural speech codec that provides state-of-the-art performance at low bitrates. We introduce and extend the back projection technique into the speech coding fields. We utilize the SDBP and SUBP modules to replace the standard and transposed convolution layers. Further, we adopt a selective feature fusion block for augmented representation. Our experiments show a significant improvement over existing methods, highlighting the effectiveness of our approach in preserving and reconstructing information for enhanced speech quality.

\label{sec:con}


\bibliographystyle{IEEEbib}
\bibliography{refs}

\end{document}